\PassOptionsToPackage{unicode}{hyperref}
\PassOptionsToPackage{hyphens}{url}
\PassOptionsToPackage{dvipsnames,svgnames,x11names}{xcolor}
\documentclass[
  12pt]{article}
  
\usepackage{listings}
\usepackage{ulem}

\usepackage{amsmath,amssymb}
\usepackage{iftex}
\ifPDFTeX
  \usepackage[T1]{fontenc}
  \usepackage[utf8]{inputenc}
  \usepackage{textcomp} % provide euro and other symbols
\else % if luatex or xetex
  \usepackage{unicode-math}
  \defaultfontfeatures{Scale=MatchLowercase}
  \defaultfontfeatures[\rmfamily]{Ligatures=TeX,Scale=1}
\fi
\usepackage{lmodern}
\ifPDFTeX\else  
\fi
\IfFileExists{upquote.sty}{\usepackage{upquote}}{}
\IfFileExists{microtype.sty}{% use microtype if available
  \usepackage[]{microtype}
  \UseMicrotypeSet[protrusion]{basicmath} % disable protrusion for tt fonts
}{}
\makeatletter
\@ifundefined{KOMAClassName}{% if non-KOMA class
  \IfFileExists{parskip.sty}{%
    \usepackage{parskip}
  }{% else
    \setlength{\parindent}{0pt}
    \setlength{\parskip}{6pt plus 2pt minus 1pt}}
}{% if KOMA class
  \KOMAoptions{parskip=half}}
\makeatother
\usepackage{xcolor}
\makeatletter
\ifx\paragraph\undefined\else
  \let\oldparagraph\paragraph
  \renewcommand{\paragraph}{
    \@ifstar
      \xxxParagraphStar
      \xxxParagraphNoStar
  }
  \newcommand{\xxxParagraphStar}[1]{\oldparagraph*{#1}\mbox{}}
  \newcommand{\xxxParagraphNoStar}[1]{\oldparagraph{#1}\mbox{}}
\fi
\ifx\subparagraph\undefined\else
  \let\oldsubparagraph\subparagraph
  \renewcommand{\subparagraph}{
    \@ifstar
      \xxxSubParagraphStar
      \xxxSubParagraphNoStar
  }
  \newcommand{\xxxSubParagraphStar}[1]{\oldsubparagraph*{#1}\mbox{}}
  \newcommand{\xxxSubParagraphNoStar}[1]{\oldsubparagraph{#1}\mbox{}}

\fi
\newcommand*{\rom}[1]{\expandafter\@slowromancap\romannumeral #1@}

\makeatother

\usepackage{longtable,booktabs,array}
\usepackage{calc} % for calculating minipage widths
\usepackage{etoolbox}
\makeatletter
\patchcmd\longtable{\par}{\if@noskipsec\mbox{}\fi\par}{}{}
\makeatother
\IfFileExists{footnotehyper.sty}{\usepackage{footnotehyper}}{\usepackage{footnote}}
\makesavenoteenv{longtable}
\usepackage{graphicx}
\makeatletter
\def\maxwidth{\ifdim\Gin@nat@width>\linewidth\linewidth\else\Gin@nat@width\fi}
\def\maxheight{\ifdim\Gin@nat@height>\textheight\textheight\else\Gin@nat@height\fi}
\makeatother
\setkeys{Gin}{width=\maxwidth,height=\maxheight,keepaspectratio}
\makeatletter
\def\fps@figure{htbp}
\makeatother

\makeatletter
\@ifpackageloaded{caption}{}{\usepackage{caption}}
\AtBeginDocument{%
\ifdefined\contentsname
  \renewcommand*\contentsname{Table of contents}
\else
  \newcommand\contentsname{Table of contents}
\fi
\ifdefined\listfigurename
  \renewcommand*\listfigurename{List of Figures}
\else
  \newcommand\listfigurename{List of Figures}
\fi
\ifdefined\listtablename
  \renewcommand*\listtablename{List of Tables}
\else
  \newcommand\listtablename{List of Tables}
\fi
\ifdefined\figurename
  \renewcommand*\figurename{Figure}
\else
  \newcommand\figurename{Figure}
\fi
\ifdefined\tablename
  \renewcommand*\tablename{Table}
\else
  \newcommand\tablename{Table}
\fi
}
\@ifpackageloaded{float}{}{\usepackage{float}}
\floatstyle{ruled}
\@ifundefined{c@chapter}{\newfloat{codelisting}{h}{lop}}{\newfloat{codelisting}{h}{lop}[chapter]}
\floatname{codelisting}{Listing}

\makeatother
\makeatletter
\@ifpackageloaded{caption}{}{\usepackage{caption}}
\@ifpackageloaded{subcaption}{}{\usepackage{subcaption}}
\makeatother

\ifLuaTeX
  \usepackage{selnolig}  % disable illegal ligatures
\fi
\usepackage[]{natbib}
\usepackage{bookmark}

\IfFileExists{xurl.sty}{\usepackage{xurl}}{} % add URL line breaks if available
\hypersetup{
  pdftitle={Title},
  pdfauthor={Author 1; Author 2},
  pdfkeywords={3 to 6 keywords, that do not appear in the title},
  colorlinks=true,
  linkcolor={blue},
  filecolor={Maroon},
  citecolor={Blue},
  urlcolor={Blue},
  pdfcreator={LaTeX via pandoc}}

\newcommand{\anon}{1}

\begin{document}

\def\spacingset#1{\renewcommand{\baselinestretch}%
{#1}\small\normalsize} \spacingset{1}

%%%%%%%%%%%%%%%%%%%%%%%%%%%%%%%%%%%%%%%%%%%%%%%%%%%%%%%%%%%%%%%%%%%%%%%%%%%%%%

\if1\anon
{
  \title{\bf Empirical Simulation of Survival and Mixed-Type Data for Clinical Trial Design}
  \author{Yao Chen \hspace{.2cm}\\
    Indiana University Indianapolis\\
    and \\
    Jiren Sun, Yuxin Ding, Yushi Liu, Yongming Qu \\
    Eli Lilly and Company}
  \maketitle
} \fi

\if0\anon
{
  \bigskip
  \bigskip
  \bigskip
  \begin{center}
    {\LARGE\bf Title}
\end{center}
  \medskip
} \fi

\bigskip
\begin{abstract}

The simulation of realistic time-to-event data is essential for the planning and evaluation of complex clinical trial designs. Traditional simulations typically draw event times from parametric families such as the Weibull and log-normal distributions, which impose a fixed hazard shape that may not reflect the true distribution of the survival data. We develop an empirical, copula-based framework that enables simulating multivariate data with mixed types of continuous, binary, count, and right-censored time-to-event variables. 
The method completes censored historical survival data through a two-zone procedure that combines conditional Kaplan--Meier imputation with a parametric tail. It matches the shape of the target survival curve through a log-scale location--scale transformation and a power distortion of the empirical percentile function. It also preserves historical dependence structure via Gaussian copula fitted to the rank correlations. We apply the method to an oncology trial in previously treated non-small-cell lung cancer; it reconstructs the overall-survival and progression-free-survival curves of the experimental arm from the control arm and a small set of target percentiles. It preserved both the rank correlations among baseline covariates and the PFS--OS rank dependence (censored Kendall's $\tau$ of $0.522$ simulated versus $0.549$ observed). The methodology is implemented in the R package \texttt{EmpiricalSim}.

\end{abstract}

\noindent%
{\it Keywords:} Gaussian copula, Location-scale family, Censored-data imputation, Quantile transformation, Synthetic patient-level data. 
\vfill

\newpage
\spacingset{1.8} % DON'T change the spacing!

\section{Introduction}\label{sec-intro}

%  The number of lines per page (letter size paper) will be about 26.

Quantitative planning of a clinical trial---power, sample size, the timing of interim analyses, and the choice of analysis method---relies on the ability to generate synthetic data that resemble the data the trial is expected to produce. In late-phase oncology trials in particular, the primary and key secondary endpoints are frequently time-to-event outcomes such as overall survival (OS) and progression-free survival (PFS), and these endpoints are correlated both with one another and with other baseline covariates such as age. In diabetes and obesity trials, many cardiometabolic biomarkers and some time-to-event outcomes like cardiovascular events are correlated. With the widespread adoption of a graphical multiple testing procedure \citep{bretz2009graphical} to control the overall family-wise Type 1 error for phase 3 clinical trials, it may be desired to evaluate the statistical power simultaneously for multivariate outcomes. In SURMOUNT-1 study \citep{jastreboff2022surmount1}, 22 comparisons were included in the graphical testing scheme. It is not possible to evaluate the statistical power under different scenarios using a closed-form analytic approach. 
A simulation framework that is useful for trial design should therefore reproduce the marginal distribution of each endpoint, and the dependence structure among endpoints and covariates, simultaneously.

The most common approach to simulating survival times is parametric. One specifies a marginal hazard family, for example, Weibull, the family that remains proportional under the Cox model \citep{cox1972regression}, and inverts the cumulative hazard to draw event times, optionally as a function of covariates \citep{bender2005generating, austin2012generating}. More flexible parametric constructions allow piecewise, spline-based, or mixture hazards to accommodate complex shapes \citep{crowther2013simulating}, and dependence among multiple event times is typically introduced through shared frailties or through copulas applied to parametric margins \citep{clayton1978model, nelsen2006introduction}. These methods are well understood, and good practice for designing and reporting the resulting simulation studies is established \citep{burton2006design, morris2019using}. Their limitation is structural rather than computational: one must commit to a marginal family, and the conclusions of the simulation study can be sensitive to that choice, because the imposed parametric distribution may not reflect the true distribution of the treatment under evaluation. % \textcolor{brown}{treatment under evaluation}

For most disease areas, individual patient--level data are available from studies previously conducted by the same sponsor, public available data, or consortia such as TransCelerate. In some cases,  
the same class of drugs is studied repeatedly across trials that differ in population, indication, and phase, and it is reasonable to expect the distribution %\textcolor{brown}{(especially the correlations) of a set of endpoints of interest} 
(especially the correlations) of a set of endpoints of interest for a given class to be broadly comparable across these studies. When historical individual patient data for the class already exist, simulating an ongoing or future study from the empirical distribution of those data---rather than from a parametric family---can more faithfully represent the distribution likely to be observed in the new study. This empirical approach additionally allows the dependence structure among covariates and endpoints estimated from the historical data to be preserved in the simulation, rather than imposed by a parametric model.

One way to use historical data directly is to resample it. Bootstrap and weighted resampling \citep{efron1979bootstrap} reproduce the historical marginal exactly, but they are locked to the historical support: they cannot extend the simulated range beyond the observed minimum and maximum, and they offer no principled mechanism for shifting the distribution to reflect an anticipated treatment effect. To realize an empirical approach that overcomes these constraints, we build on the framework of \citet{ding2026empirical}, who proposed a hybrid method that retains the non-parametric fidelity of resampling while regaining the flexibility of a location--scale family. Their construction maps historical data to a standard normal space through the probability integral transform, preserves the rank-correlation structure with a Gaussian copula, samples new latent normal vectors, and maps them back through a power-distortion of the uniform quantiles, $u^{\alpha}(1-u)^{\beta}$, whose two parameters are tuned by grid search to match a target mean and standard deviation. A location--scale shift $(\tau,\delta)$ then permits clinically plausible extrapolation beyond the historical range. The framework handles continuous, binary, and count endpoints, but it does not address survival data, for three connected reasons. First, right-censoring makes the empirical cumulative distribution function unreliable, so the rank-based inverse $\hat{F}^{-1}(\cdot)$ that the method depends on is biased downward if censored observations are simply dropped. Second, a linear location--scale shift can map a positive survival time to a negative value, which is inadmissible. Third, the mean and standard deviation are poorly estimated and poorly interpretable under heavy censoring, so they are unsuitable as matching targets.

This paper develops an extension of the \citet{ding2026empirical} framework that brings empirical, assumption-lean simulation to censored survival endpoints and to the joint simulation of survival and mixed-type covariates. We add an imputation to handle the censored historical data using the Kaplan--Meier estimator \citep{kaplan1958nonparametric} and a working parametric model. We match targeted survival times with Kaplan--Meier percentiles, and adapt the power-distortion grid search to match the shape of survival curve. Throughout, dependence is carried by the Gaussian copula, so the joint structure between survival endpoints and covariates is preserved by construction. Section~\ref{sec-meth} describes the method and connects each component to its implementation in the R package. Section~\ref{sec-app} applies the method to the OAK trial \citep{rittmeyer2017atezolizumab}. Section~\ref{sec-disc} discusses the results, the relationship to existing methods, and limitations.

\section{Methods}\label{sec-meth}

\subsection{Notation and inputs}

Let the historical arm consist of $n_0$ patients. For a survival endpoint, patient $i$ contributes an observed time $X_i=\min(T_i,C_i)$ and an event indicator $\Delta_i=\mathbb{I}(T_i\le C_i)$, where $T_i$ is the true event time and $C_i$ the censoring time; a study may contain more than one such endpoint (for example PFS and OS). In addition to survival endpoints, each patient has a vector of $p$ covariates that may be continuous, binary, or count. The method specifies the joint distribution of the survival endpoints and covariates through its marginals and a copula, calibrating each marginal separately while modeling their dependence through the copula of Section~\ref{sec:method:transform}.

For each survival endpoint the user supplies a small table of target survival percentiles (pairs $(q_k, v_k)$ stating that a fraction $q_k$ of patients on the new treatment are expected to have experienced the event by time $v_k$) and, for each non-survival covariate, target summaries (mean, standard deviation, and a lower and upper bound to anchor the range).  No patient-level data for the new arm are needed. The goal is to generate $N$ synthetic patients whose marginal distributions are calibrated to these targets, whose support is clinically plausible, and whose joint dependence mimics the historical data.

We generate a continuous, binary, or count endpoint by mapping a latent Gaussian draw through the copula, distorting the resulting uniform by $h(u)=u^{\alpha}(1-u)^{\beta}$, passing it through the empirical quantile function $\hat{F}^{-1}$, and applying a linear location--scale shift. For survival endpoint, we replace the linear shift with a log-scale one, so that simulated times remain strictly positive, and feeds $\hat{F}^{-1}$ from a censoring-completed sample (Section~\ref{sec-impute}). Writing $u^{*}=\Phi(z^{*})$ 
for the copula uniform of a single endpoint, the two constructions are
\begin{align}
  y^{*} &= \delta + \tau\,\hat{F}^{-1}\!\big(h(u^{*})\big) , \label{eq-ding}\\[2pt]
  y^{*} &= \exp\!\Big\{\delta + \tau\,\log \hat{F}^{-1}\!\big(h(u^{*})\big)\Big\},
        \label{eq-master}
\end{align}
where the first is the mixed-type construction for continuous, binary, and count endpoints and the second is the survival construction introduced here. They only differ in the link (identity versus log) and in how $(\tau,\delta)$ are calibrated---to a target mean and standard deviation in the former Eq.~\eqref{eq-ding}, to target survival percentiles in the latter Eq.~\eqref{eq-master} (Section~\ref{sec-shape}).

Equation~\eqref{eq-master} is written for one survival endpoint, but the latent draw is shared across all of them. Let $\mathbf{z}^{*}\sim N(0,\widehat{\Omega})$ be a single Gaussian vector drawn once per synthetic patient, whose covariance $\widehat{\Omega}$ encodes the historical rank correlations among all variables: every survival endpoint and every mixed-type covariate. 

Each margin is generated from its own coordinate of $\mathbf{z}^{*}$: the $z^{*}$ in Eq.~\eqref{eq-master} is the coordinate belonging to the survival endpoint being generated, and it is the sharing of $\mathbf{z}^{*}$ across coordinates that carries the joint dependence (Section~\ref{sec-pipeline}). In the remaining terms, $\Phi$ is the standard normal cumulative distribution function, $\hat{F}^{-1}$ is the empirical quantile function of the (imputed) historical event times, $(\alpha,\beta)$ are shape parameters that distort the uniform quantile, and $(\tau,\delta)$ are the log-scale scale and location parameters.

 \subsection{Imputation of right-censored observations}\label{sec-impute}

Because the simulated times in Eq.~\eqref{eq-master} are drawn from the sorted historical event times through $\hat{F}^{-1}$, censored observations cannot simply be discarded or retained at their censoring
times. Doing so biases the empirical quantile function. We therefore impute the censored observations before any resampling. The imputation partitions the sorted historical event times at a truncation point, and treats the two regions differently.

The truncation point is determined by Peto's effective sample size \citep{peto1972asymptotically}. Specifically,
\begin{equation}\label{eq-neff}
    n_{\mathrm{eff}}(t)=\frac{\hat{S}(t)\big(1-\hat{S}(t)\big)}
    {\widehat{\mathrm{Var}}[\hat{S}(t)]},
\end{equation}
where the variance $\widehat{\mathrm{Var}}[\hat{S}(t)]$ is estimated using Greenwood’s formula \citep{greenwood1926natural}. The truncation time $t_{\mathrm{trunc}}$ is defined as the largest event time for which $n_{\mathrm{eff}}(t)$ exceeds a user-specified threshold (default 10). 

For a censored observation with last follow-up time $c<t_{\mathrm{trunc}}$ (``Zone~1''), the Kaplan--Meier estimate is regarded as sufficiently reliable because the effective sample size exceeds the truncation threshold. An imputed event time is drawn from the conditional Kaplan--Meier distribution given survival beyond $c$, by sampling $u\sim\mathrm{Unif}(0,\hat{S}(c))$ on the survival-probability scale and setting the imputed event time to $\inf\{t>c:\hat S(t)\le u\}$.  For a censored observation with $c\ge t_{\mathrm{trunc}}$ (``Zone~2''), where the Kaplan--Meier estimate is unstable because of the small effective sample size, the replacement is drawn from a fitted parametric tail.

The parametric tail is selected once, on the full data, from five candidate families (Weibull, log-normal, log-logistic, generalized gamma, Gompertz). Because the fitted model is used only to extrapolate the upper tail (Zone~2), each family is estimated by \emph{weighted} maximum likelihood rather than by ordinary maximum likelihood, up-weighting the log-likelihood contributions of events occurring after a tail-start time $t_{\mathrm{tail}}$ (a fraction of $t_{\mathrm{trunc}}$, default one half) by a multiplier (default three). With per-observation weights $w_i$, each family is fitted by maximizing the weighted log-likelihood, $\hat{\theta}=\arg\max_{\theta}\sum_{i=1}^{n_0} w_i\,\ell_i(\theta)$, where $\ell_i(\theta)$ is the log-likelihood contribution of observation $i$ and the $w_i$ are supplied as case weights \citep{jackson2016flexsurv}. Weighted maximized log-likelihood $\sum_i w_i\,\ell_i(\hat{\theta})$ is used for model comparison, and the candidate families are ranked by the corresponding AIC,
\begin{equation}\label{eq-twaic}
  \mathrm{AIC}_{\mathrm{tw}} \;=\; -2\sum_{i=1}^{n_0} w_i\,\ell_i(\hat{\theta}) + 2k,
  \qquad
  w_i = \begin{cases} 3, & X_i \ge t_{\mathrm{tail}}, \\ 1, & \text{otherwise,}
  \end{cases}
\end{equation}
with $k$ the number of parameters. Comparing models fitted by weighted likelihood on a weighted information criterion follows \citet{konishi1996generalised}, and the case-weighted form specifically follows \citet{lumley2015aic}. Fitting each family by weighted maximum likelihood concentrates its accuracy in the tail rather than in the densely observed early region, aligning the estimate with the region the selection criterion evaluates; the family chosen is therefore the one that best describes the part of the distribution the model is actually asked to extrapolate.

Imputation can be repeated $M$ times with different seeds to support multiple imputation \citep{rubin1987multiple}; the diagnostics returned include $t_{\mathrm{trunc}}$, $\hat{S}(t_{\mathrm{trunc}})$, the tail-weighted AIC table, and the counts of Kaplan--Meier- versus parametric-imputed observations. After this step the historical arm is a complete (uncensored) sample, and all event indicators are set to one for the purpose of building $\hat{F}^{-1}$.

The imputation is a device that lets the framework operate on a complete sample; it is not a claim to recover the unobserved event times individually. In a simulated trial with administrative censoring, the same censoring mechanism is applied to the synthetic patients, so imputed times that fall beyond the study's follow-up horizon are re-censored and never enter the analysis. The imputed values therefore matter only insofar as they shape $\hat{F}^{-1}$ over the range the simulated study can actually observe; under heavy administrative censoring, most of them lie past the cutoff and are discarded, and the method's behavior in the observable range is governed largely by the observed Kaplan–Meier curve rather than by the parametric tail. 
%\textcolor{red}{I don't think it is purely by K-M. Maybe we should say "largely by the obsereved K-M curve"} 
This is what distinguishes the approach from a fully parametric simulation, whose assumed family drives the fit throughout, including the densely observed early region.

\subsection{Copula transform and correlation preservation}\label{sec:method:transform}

Dependence is preserved exactly as in \citet{ding2026empirical}. Each historical variable is mapped to a uniform score through its (mid-)ranks, $F_u = (\mathrm{rank}(y) - 0.5)/n_0$, and then to the normal scale through $\Phi^{-1}(F_u)$. The sample covariance of these normal scores is the copula correlation matrix $\widehat{\Omega}$. New latent vectors are drawn as $z^{*}\sim N(0,\widehat{\Omega})$, and transformed to the copula uniforms $u=\Phi(z^{*})$.  Exact preservation requires the marginal map to be monotone. This holds for the log-percentile scaling and for the power distortion when $\beta=0$; for $\beta>0$, $h(u)=u^{\alpha}(1-u)^{\beta}$ is non-monotone for $u> u^*=\alpha/(\alpha+\beta)$, so rank preservation in the extreme upper tail is approximate, with the error governed by the small probability mass in that region.
 
\subsection{Log-percentile location--scale transformation}

Applying the linear shift $y^{*}=y\tau+\delta$ used for continuous endpoints directly to time-to-event data can yield negative values and is therefore inadmissible for survival times. We instead work on the log scale. Two of the target percentiles serve as scaling anchors: the Scale-Low and the Scale-High percentiles. Writing $v_{\mathrm{SL}}^{*}, v_{\mathrm{SH}}^{*}$ for the target times at these levels and $v_{\mathrm{SL}}, v_{\mathrm{SH}}$ for the corresponding times of the historical arm, the log-scale parameters are

\begin{equation}\label{eq-taudelta}
  \tau \;=\; \frac{\log v_{\mathrm{SH}}^{*} - \log v_{\mathrm{SL}}^{*}}
                  {\log v_{\mathrm{SH}} - \log v_{\mathrm{SL}}},
  \qquad
  \delta \;=\; \log v_{\mathrm{SL}}^{*} - \tau\,\log v_{\mathrm{SL}}.
\end{equation}

The map $y\mapsto\exp(\tau\log y + \delta)$ is monotone increasing on the positive reals and returns strictly positive times. We use Kaplan--Meier percentiles as anchors rather than the minimum and maximum survival times, and by default the quantile range is drawn from the portion of the historical survival curve still supported by observed data, excluding the extrapolated tail where the curve relies on imputation alone.
  
\subsection{Shape matching by power distortion}\label{sec-shape} 

With $\widehat{\Omega}$ in hand, the simulation draws $z^{*}\sim N(0,\widehat{\Omega})$ and forms $u=\Phi(z^{*})\sim\mathrm{Unif}(0,1)$ marginally. For a univariate survival simulation, $\widehat{\Omega}$ reduces to a scalar and $z^{*}$ is a standard normal draw; for the joint simulation it carries the full cross-endpoint correlation.

The uniform $u$ is distorted by $h(u)=u^{\alpha}(1-u)^{\beta}$ before being passed through $\hat{F}^{-1}$. The parameter $\alpha$ stretches or compresses the body of the distribution and $\beta$ adjusts the upper tail. The pair $(\alpha,\beta)$ is chosen to match three interior target percentiles---Target-Low ($q_{\mathrm{low}}$), Target-Med ($q_{\mathrm{med}}$), and Target-High ($q_{\mathrm{high}}$)---of the \emph{transformed} historical distribution to their target times, where the transformation includes the log-percentile scaling of Eq.~\eqref{eq-taudelta}. The search is a vectorized two-stage grid:

\begin{description}

\item[Stage 1 (coarse $\alpha$ scan).] With $\beta=0$, sweep $\alpha\in[0.2,2.0]$ on a grid of step $0.01$ and select $\alpha^{*}=\arg\min_{\alpha}|\Delta_{\mathrm{med}}|$, where $\Delta_{\mathrm{med}}$ is the discrepancy between the simulated and target median. Record the residual $\epsilon_1=|\Delta_{\mathrm{med}}|$.

\item[Stage 2 (refined $(\alpha,\beta)$ search).] Restrict $\alpha$ to a narrow window around $\alpha^{*}$, refine the step to $0.001$, and add $\beta\in[0,\beta_{\max}=0.01\alpha]$. Retain only those $(\alpha,\beta)$ pairs whose median discrepancy stays within a relaxation multiple of $\epsilon_1$, and among those select the pair minimizing $|\Delta_{\mathrm{low}}|+|\Delta_{\mathrm{high}}|$, the combined error at the lower and upper interior percentiles. The upper bound $\beta_{\max}$ is kept small because, as noted in Section~\ref{sec:method:transform}, $h(u)=u^{\alpha}(1-u)^{\beta}$ becomes non-monotone for $u>\alpha/(\alpha+\beta)$ once $\beta>0$; a small $\beta$ confines this non-monotonicity to the extreme upper tail, where little probability mass lies, so that the distortion needed to bend the tail toward its target is achieved with only negligible violation of rank preservation. 
\end{description} 

Because the turning point is $u^{*}=\alpha/(\alpha+\beta)$, an absolute cap on $\beta$ does not confine the non-monotone region consistently across the $\alpha$-range, for example, as in Table~\ref{tbl-beta-max}, larger values move the non-monotone region inward and can materially distort the simulated distribution. We therefore cap $\beta$ relative to $\alpha$ instead of by an absolute constant,  $\beta_{\max}=0.01\,\alpha$, which fixes $u^{*}\ge 1/1.01\approx0.99$  regardless of the value of $\alpha^{*}$ selected in Stage~1. The non-monotone region stays confined above the $99$th percentile of $u$ for every $\alpha$ considered, rather than drifting inward when $\alpha^{*}$ happens to be small. 

\begin{table}[htbp]
\centering
\caption{Size of the non-monotone region of the power distortion $h(u)=u^{\alpha}(1-u)^{\beta}$, reported as the upper-tail probability mass $1-u^{*}=\beta/(\alpha+\beta)$ that lies above the turning point $u^{*}$, under three caps on $\beta$. Absolute caps ($\beta_{\max}=0.1$ or $0.05$) let this non-monotone region grow large when $\alpha$ is small, so that non-monotonicity intrudes well into the body of the distribution. The relative cap $\beta_{\max}=0.01\,\alpha$ instead holds it at $1.0\%$ for every $\alpha$, confining non-monotonicity above the $99$th percentile of $u$.}
\label{tbl-beta-max}
\begin{tabular}{rrrr}
\toprule 
$\alpha$ & $\beta_{\max} = 0.1$ & $\beta_{\max} = 0.05$ & $\beta_{\max} = 0.01\alpha$ \\
\midrule
$0.2$ & $33.3\%$ & $20.0\%$ & $1.0\%$ \\
$1.0$ & $9.1\%$ & $4.8\%$ & $1.0\%$ \\
$1.5$ & $6.3\%$ & $3.2\%$ &  $1.0\%$ \\
$2.0$ & $4.8\%$ & $2.4\%$ &  $1.0\%$ \\
\bottomrule
\end{tabular}
\end{table}

The three-target scheme above is the default rather than a requirement. When only a central target is available or of interest, the search can reduce to Stage~1 alone, matching $q_{\mathrm{med}}$ with $\beta=0$; when more than three interior percentiles are supplied, Stage~2 generalizes by minimizing the summed discrepancy over all targets other than $q_{\mathrm{med}}$, subject to the same relaxation. The number of interior targets is therefore chosen by the user according to how much is known about the intended distribution.

The same routine matches mean and standard deviation for continuous, binary, and ordinal endpoints, the only differences being the matching objective and a linear rather than log-scale shift; this unifies survival and mixed-type endpoints under a single optimizer.

\subsection{Choosing the target percentiles}\label{sec-percentiles}

The percentile--time pairs are the main input controlling a simulated survival endpoint, and they play two separate roles. The two extreme percentiles, Scale-Low and Scale-High, set the \emph{range}: through Eq.~\eqref{eq-taudelta} they determine how far the simulated distribution is shifted and stretched relative to the historical one. The three interior percentiles, Target-Low, Target-Med, and Target-High, set the \emph{shape}: the power-distortion search of Section~\ref{sec-shape} bends the historical curve until it passes through them. Choosing percentiles therefore comes down to two questions: what range to cover, and where inside that range to pin the shape. How the two are balanced depends on how the new arm is expected to differ from the historical control.

By default we simply read the levels off the historical curve. We use only the part of the curve that survives the effective-sample-size truncation of Section~\ref{sec-impute}, beyond which the Kaplan--Meier estimate becomes too noisy to anchor to. Within that interval we place five levels: the outer two (near the $10$th and $90$th event percentiles of the control) as the scaling anchors, and the inner three (near the $30$th, $50$th, and $70$th) as the shape targets. When censoring shortens the reliable interval, all five levels contract with it so that they stay inside it. Suppose, for example, that the historical overall-survival curve is available only up to its $70$th event percentile---the time by which $70\%$ of patients have had the event and only $30\%$ are still surviving. The five levels are then compressed into this shorter window instead of spanning the full $10$th-to-$90$th event percentile, with the two scaling anchors falling at about the $7$th and $63$rd event percentiles and the three shape targets spaced between them.

When the analyst has prior knowledge of how the new arm should differ, the levels can instead be set manually: the interior anchors are moved toward the part of the curve where the difference is expected. Because the shape of the new arm is usually unknown in advance, most applications begin from the defaults and depart from them only to reflect a genuine clinical expectation.

\subsection{Pipeline}\label{sec-pipeline}

The complete procedure combines the preceding components into a single pass that produces every margin from one correlated latent draw. Given the historical arm and the user's target summaries, the simulation proceeds in six steps.

\begin{enumerate}
\item \emph{Impute censoring.} For each survival endpoint, complete the right-censored historical times by the two-zone procedure of
Section~\ref{sec-impute}, obtaining an uncensored sample and the empirical quantile function $\hat{F}^{-1}$ over the sorted imputed times.
\item \emph{Fit the copula.} Map every historical variable---survival endpoints and mixed-type covariates---to normal scores, and estimate the copula correlation matrix $\widehat{\Omega}$ (Section~\ref{sec:method:transform}).
\item \emph{Calibrate each margin.} For each survival endpoint, solve for the log-scale parameters $(\tau,\delta)$ from the two scaling anchors (Eq.~\eqref{eq-taudelta}) and for the shape parameters $(\alpha,\beta)$ by the two-stage grid search against the interior targets (Section~\ref{sec-shape}); for each continuous, binary, or ordinal covariate, run the same optimizer with a linear shift and a mean/standard-deviation objective. 
\item \emph{Draw the shared latent sample.} Sample $N$ vectors $z^{*}\sim N(0,\widehat{\Omega})$, one per synthetic patient, and form the copula uniforms $u=\Phi(z^{*})$.
\item \emph{Back-transform every column from the same $z^{*}$.} Apply each margin's calibrated map to its own coordinate of $u$: Eq.~\eqref{eq-master} for survival endpoints, empirical $\hat{F}^{-1}$ lookup with the linear shift for continuous, binary and ordinal covariates. Because all columns are generated from the one latent sample, the marginal calibration of Step~3 and the joint dependence of $\widehat{\Omega}$ are achieved simultaneously.
\end{enumerate}

The pipeline is implemented in the \texttt{EmpiricalSim} R package via \texttt{run\_simulation}, which takes the historical data, the per-variable target tables, and the sample size and returns the simulated data set together with per-step diagnostics---the imputation summary, the fitted $(\tau,\delta)$ and $(\alpha,\beta)$, and the realized-versus-target percentiles---so that each margin and the dependence structure can be checked before use. 
\section{Application to the OAK Trial}\label{sec-app}

\subsection{Trial and objective}
 
The OAK study was an open-label, $1{:}1$ randomized, multinational phase~\rom{3} trial in previously treated non-small-cell lung cancer that compared atezolizumab with docetaxel \citep{rittmeyer2017atezolizumab}. Our application uses the tumor mutational burden (TMB)--evaluable population, a subset of the intention-to-treat (ITT) population comprising the OAK patients with an adequate plasma sample for TMB assessment \citep{gandara2018blood}. This biomarker-evaluable population consisted of $290$ patients in the docetaxel arm and $293$ in the atezolizumab arm. For overall survival, there were $233$ events and $57$ censored observations in the docetaxel arm, and $195$ events and $98$ censored observations in the atezolizumab arm; for progression-free survival, there were $270$ events and $20$ censored observations in the docetaxel arm, and $261$ events and $32$ censored observations in the atezolizumab arm. We treated the docetaxel arm as the available historical control and asked the method to simulate a new treatment arm from target percentiles alone, then compared the simulated arm with the actual atezolizumab arm, which was held out and used only for validation.

\subsection{Results}

 \paragraph*{Target percentiles.} 

\begin{figure}[htbp]
\centering
\includegraphics[width=\linewidth]{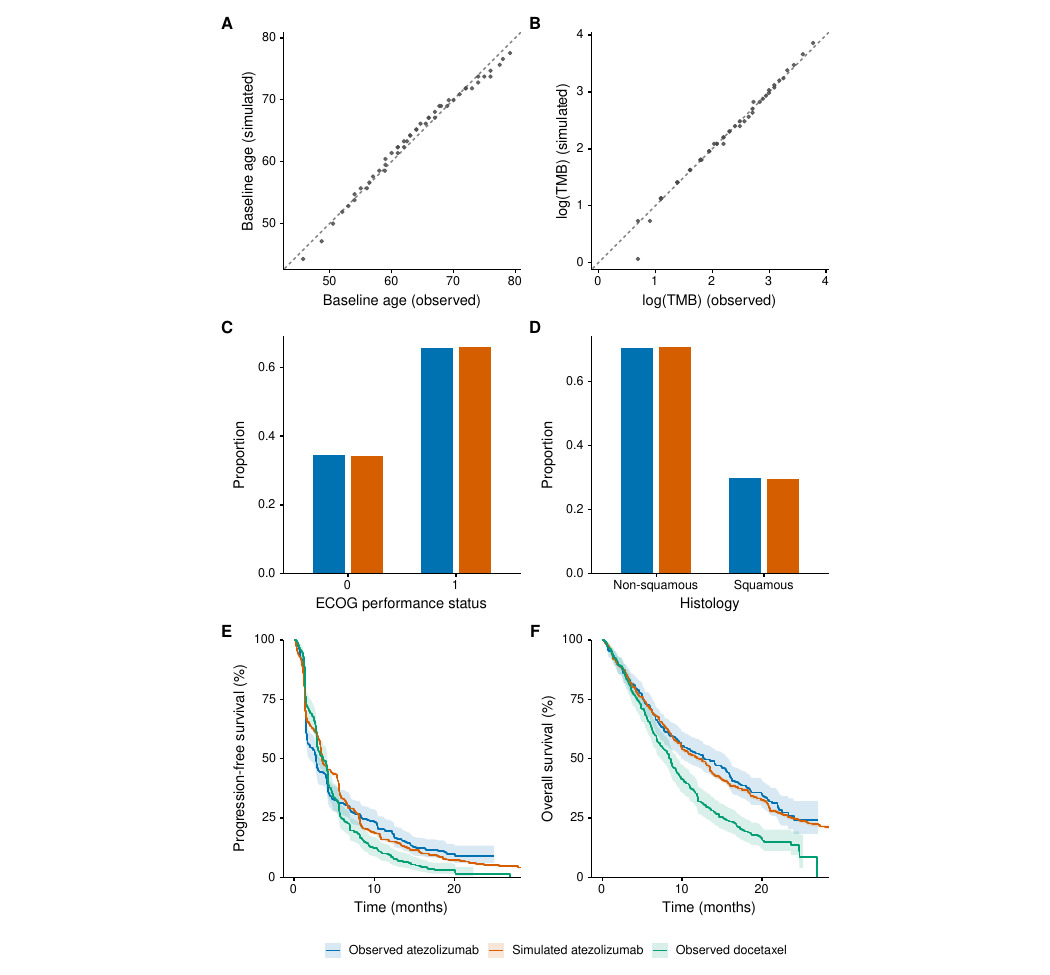}
\caption{Marginal fidelity in the OAK application. Quantile--quantile plots of
simulated versus observed (A) baseline age and (B) log tumor mutational burden;
simulated and observed proportions of (C) ECOG performance status and (D) histology; and Kaplan--Meier curves for (E) progression-free survival and (F) overall survival, with simulated curves overlaid on the observed curves and their pointwise confidence bands. The overall-survival arm was reconstructed from five target percentiles and the docetaxel control history alone; the observed atezolizumab curve was held out for validation.}
\label{fig-marginal}
\end{figure}

Because the atezolizumab arm was held out, we set the target times to its Kaplan--Meier percentiles, so that we can test whether the method reconstructs an entire curve and its dependence structure from only five points on it. The two endpoints required different placements. Overall survival curves have a similar shape, and the experimental atezolizumab arm (Blue line in Figures~\ref{fig-marginal}F) behaves like a shifted, rescaled docetaxel control (Green line in Figures~\ref{fig-marginal}F), so the decisive choice was the range and in particular how far the upper anchor reached into the tail where the benefit accrues. Within the reliably KM estimable portion of the docetaxel curve, which censoring limits to about its $76$th event percentile, we placed five overall-survival levels---Scale-Low ($8\%$ at $1.7$ months), Target-Low ($23\%$ at $5.0$), Target-Med ($38\%$ at $7.8$), Target-High ($53\%$ at $14.9$), and Scale-High ($68\%$ at $20.9$)---with the outer two as scaling anchors and the inner three as shape targets.

Progression-free survival instead differs in shape: the two curves cross (Blue and Green line in Figures~\ref{fig-marginal}E) , docetaxel higher early and atezolizumab higher later, so the three interior targets were positioned to bracket the crossover, concentrating the shape search where the distributions diverge, while the anchors fixed the ends of the range.

In a prospective design the experimental-arm shape would be unknown, so these levels would instead be read from the historical curve by default (Section~\ref{sec-percentiles}) and shifted only to reflect clinical expectations.

\paragraph*{Simulation run.}
The docetaxel survival data were first completed by the imputation step of Section~\ref{sec-impute}. The log-percentile scaling of Eq.~\eqref{eq-taudelta} and the two-stage power search of Section~\ref{sec-shape} were then applied to yield a single set of parameters, from which $10{,}000$ synthetic patients were drawn. The procedure was run jointly for PFS, OS, and the four baseline covariates (age, log tumor mutational burden, ECOG performance status, and histology) through \texttt{run\_simulation} in the \texttt{EmpiricalSim} package, so that the cross-endpoint and endpoint--covariate dependence was carried by the shared Gaussian copula. Throughout, the atezolizumab arm was held out and used only for validation. 

\begin{figure}[htbp]
\centering
\includegraphics[width=\linewidth]{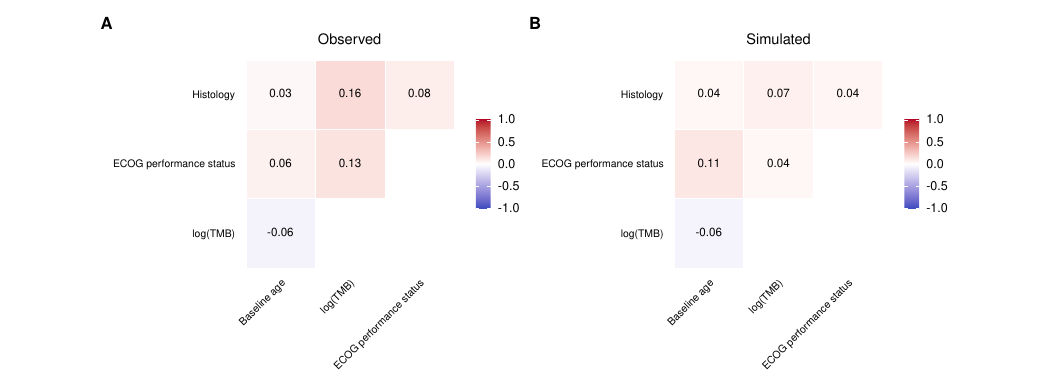}
\caption{Rank-correlation structure of the baseline covariates in the OAK application: (A) observed, and (B) simulated. The simulated matrix recovers the sign and magnitude of each observed association; the largest absolute difference is $0.09$.}
\label{fig-corr}
\end{figure}

\begin{figure}[htbp]
\centering
\includegraphics[width=0.85\linewidth]{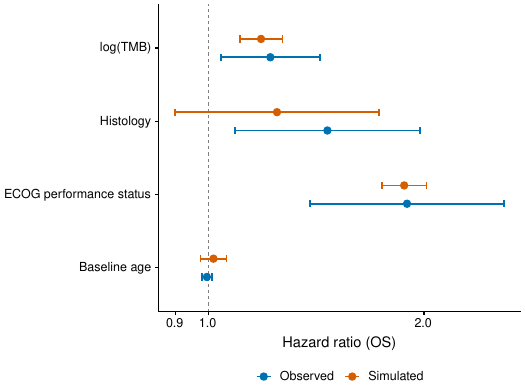}
\caption{Univariate Cox hazard ratios for overall survival, estimated on the observed
and simulated data, with $95\%$ confidence intervals. For every covariate the simulated
estimate falls within the observed confidence interval.}
\label{fig-cox}
\end{figure}

The four baseline covariates were chosen both to span the supported data types and to reflect established factors in this disease: baseline age and log tumor mutational burden (TMB) are continuous, ECOG performance status is ordinal (values $0$ and $1$ in this population), and histology (squamous versus non-squamous) is binary. Tumor mutational burden is a measure of tumor neoantigen load and a predictive biomarker for response to immune-checkpoint inhibition: tumors carrying more mutations present more neoantigens and are more likely to be recognized by the immune system, and higher TMB has been associated with greater benefit from anti-PD-L1 therapy in non-small-cell lung cancer \citep{gandara2018blood, han2026threshold}. We model TMB on the log scale because its distribution is strongly right-skewed. For a continuous covariate such as log TMB, generation reduces to an empirical-percentile back-transformation of the shared latent draw (Section~\ref{sec-meth}), with the Gaussian copula carrying its correlations with the remaining covariates and with PFS and OS; the fidelity of this reconstruction is shown in Figures~\ref{fig-marginal}B, \ref{fig-corr}, and \ref{fig-cox}.

\paragraph*{Marginal fidelity.}
Figure~\ref{fig-marginal} compares the simulated and observed margins for every endpoint and covariate. The quantile--quantile plots for baseline age (Figure~\ref{fig-marginal}A) and log tumor mutational burden (Figure~\ref{fig-marginal}B) lie along the identity line across their full range, and the simulated proportions of ECOG performance status (Figure~\ref{fig-marginal}C) and squamous versus non-squamous histology (Figure~\ref{fig-marginal}D) reproduce the observed proportions. The two time-to-event endpoints are recovered as well: the simulated progression-free-survival curve (Figure~\ref{fig-marginal}E) and overall-survival curve (Figure~\ref{fig-marginal}F) track the observed curves across the entire follow-up and remain within the pointwise $95\%$ confidence band, reproducing both the early decline and the long upper tail that a single parametric family would struggle to capture simultaneously. Because the simulated-experimental-arm curve was reconstructed from five target percentiles and the control history rather than estimated from observed experimental-arm data, this comparison shows that our method recovers the full shape of a non-proportional curve from a small number of anchors.

Table~\ref{tbl-oak} quantifies the overall-survival and progression-free survival probabilities at five reporting percentiles. The simulated and held-out atezolizumab survival times agree to within roughly one to two months, with the largest discrepancy confined to the extreme upper tail.

\begin{table}[htbp]
\centering
\caption{OAK application: Kaplan--Meier survival percentiles (months) for the simulated immunotherapy arm versus the held-out atezolizumab arm, reported for overall survival (OS) and progression-free survival (PFS). Each simulated arm was generated from the docetaxel control and five target percentiles only.}
\label{tbl-oak}
\begin{tabular}{rrrr}
\toprule
Percentile (\%) & Simulated & Observed & Difference \\
\midrule
\multicolumn{4}{l}{\emph{Overall survival (months)}} \\
$92$ & $1.3$  & $1.7$  & $-0.4$           \\
$77$ & $4.5$  & $5.0$  & $-0.5$           \\
$62$ & $8.6$  & $7.8$  & $\phantom{-}0.8$ \\
$47$ & $13.5$ & $14.9$ & $-1.4$           \\
$32$ & $20.6$ & $20.9$ & $-0.3$           \\
\addlinespace
\multicolumn{4}{l}{\emph{Progression-free survival (months)}} \\
$80$ & $1.2$  & $1.3$  & $-0.1$           \\
$70$ & $1.3$  & $1.4$  & $-0.1$           \\
$60$ & $1.5$  & $2.5$  & $-1.0$           \\
$35$ & $5.6$  & $4.7$  & $\phantom{-}0.9$ \\
$9$  & $16.5$ & $15.1$ & $\phantom{-}1.4$ \\
\bottomrule
\end{tabular}
\end{table}

\paragraph*{Covariate dependence.}
Marginal fidelity alone may not be sufficient for trial design; the joint structure should also be reproduced. Figure~\ref{fig-corr} shows the rank-correlation matrix of the baseline covariates in the observed data (panel 2A), and the simulated data (panel 2B). The simulated correlations recover the sign and approximate magnitude of every observed pairwise association, and the element-wise differences are small, with the largest absolute discrepancy at $0.09$. The dependence among baseline covariates is therefore carried through the Gaussian copula rather than destroyed by the marginal shape matching.
 
\paragraph*{Covariate effects on survival.}
Because the simulated covariates retain their association with the survival endpoints, covariate--outcome relationships estimated on the synthetic data match those in the observed data. Figure~\ref{fig-cox} reports univariate Cox hazard ratios for overall survival for each baseline covariate. For all four covariates the simulated point estimate lies within the confidence interval of the corresponding observed hazard ratio, and the two sets of intervals overlap substantially, indicating that an analysis calibrated on the simulated arm would draw the same qualitative conclusions about effects as one calibrated on the real arm.
 
 \begin{figure}[htbp]
\centering
\includegraphics[width=\linewidth]{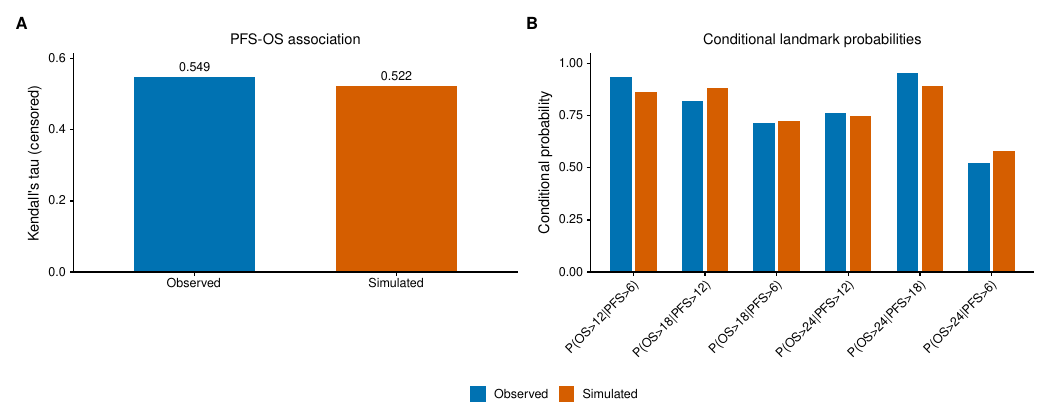}
\caption{Joint PFS--OS dependence in the OAK application. (A) Censored Kendall's $\tau$ between PFS and OS, observed versus simulated. (B) Conditional landmark probabilities $P(\mathrm{OS}>s \mid \mathrm{PFS}>r)$ for six $(r,s)$ combinations, observed versus simulated. The simulated data reproduce both the global rank association and the conditional landmark probabilities.}
\label{fig-joint}
\end{figure}
\paragraph*{Cross-endpoint dependence.}
The simulator can not only reproduce the marginal curves of PFS and OS, but also preserve the dependence between PFS and OS. Figure~\ref{fig-joint}A shows that the censored Kendall's $\tau$ between PFS and OS is $0.52$ in the simulated data against $0.55$ in the observed data. Figure~\ref{fig-joint}B compares six conditional landmark probabilities of the form $P(\mathrm{OS}>s \mid \mathrm{PFS}>r)$; the simulated and observed probabilities agree closely across all six conditions. The joint PFS--OS structure is thus preserved, so the simulated data would correctly inform the
design of any analysis that borrows strength across the two endpoints.

\section{Discussion}\label{sec-disc} 
%\textcolor{brown}{We proposed a method to simulate clinical trial data based on historical individual patient--level data for multivariate outcomes of mixed type of continuous, binary, count, and time-to-event variables. This allows a systematic and simultaneous evaluation of design properties for multiple efficacy and safety endpoints. This research extends} \sout{We have extended} 

We proposed a method to simulate clinical trial data based on historical individual patient--level data for multivariate outcomes of mixed type of continuous, binary, count, and time-to-event variables. This allows a systematic and simultaneous evaluation of design properties for multiple efficacy and safety endpoints. This research extends the empirical, copula-based simulation framework of \citet{ding2026empirical} to right-censored survival endpoints and to the joint simulation of survival outcomes with mixed-type covariates. 
%\sout{baseline} covariates. \textcolor{red}{why only baseline covariates?} 
The extension rests on three components: (1)  a two-zone imputation that completes censored historical data using the Kaplan--Meier distribution and a tail-weighted working parametric fit; (2)  a log-scale location--scale transformation; and  (3) a power distortion of the empirical percentile function that matches the shape of a target survival curve from as few as three percentile--time targets. Dependence is induced by a Gaussian copula fitted  to the historical rank correlations, therefore the joint structure among endpoints and covariates is preserved by construction rather than imposed through a parametric
model.
 
Fully parametric survival simulation \citep{bender2005generating, austin2012generating, crowther2013simulating} is flexible and well documented \citep{burton2006design, morris2019using}, but it requires committing to a marginal hazard family; when the historical curve combines an early steep decline with a long plateau, no single low-dimensional family captures both regions. Direct resampling \citep{efron1979bootstrap} avoids that commitment but is confined to the observed support and does not shift the distribution toward an anticipated treatment effect. Our approach keeps the assumption-free fidelity of resampling in the well-observed body of the distribution, while adding the ability---which resampling lacks---to shift and reshape it into the distribution expected for the new treatment arm, using only a set of target summaries.
 
A difficulty in survival summarization is the unreliability of the extreme tail of the Kaplan--Meier estimate, where few patients remain at risk. The same difficulty motivates the choice of a threshold (truncation) time in restricted mean survival time (RMST) analysis, for which data-driven selection procedures based on hazard-rate changepoints have recently been proposed  \citep{han2026threshold}. Our truncation time $t_{\mathrm{trunc}}$, determined by an effective-sample-size criterion, plays an analogous role: it separates the region in which the empirical curve is trusted from the region governed by a working parametric tail. Because RMST is a natural non-proportional-hazards summary for the trials this method is designed to simulate, data generated by the framework are well suited to planning analyses that compare arms on the restricted-mean scale and to studying how such comparisons depend on the chosen threshold time.

The framework suits settings in which one needs individual-level data for a trial arm that is unobserved or only partially observed, drawing on a historical distribution of comparable shape. Several recurring situations arise. First, an early-phase trial can supply the historical arm for a later-phase design: phase~\rom{2} patient data can be expanded to a phase~\rom{3}-sized sample that preserves the observed covariate--endpoint dependence, while the location--scale and power-distortion adjustments shift and reshape the empirical curve to reflect the anticipated phase~\rom{3} treatment effect, supporting probability-of-study-success calculations and the planning of graphical testing. Second, when a design calls for a percentile--time relationship unlike the historical one, the interior targets bend the simulated distribution into a delayed-effect, crossing, or otherwise non-proportional shape---as in Section~\ref{sec-app}---without committing to a parametric hazard family. Third, when only a published report is available, targets can be read from the reported Kaplan--Meier plots and used to simulate a patient--level data set for ad hoc analyses the original did not report, such as restricted mean survival time under an alternative threshold, to support competitive landscape analysis. Fourth, when the patient-level data that would illustrate a new statistical method cannot be shared for confidentiality reasons, the method can generate a synthetic data set calibrated to those data to serve as the worked example in the publication.

We close with several considerations. First, the method assumes that the endpoint distribution of the new arm is similar in shape to that of the historical data, up to the location--scale and power-distortion adjustments encoded by the target percentiles. This is the premise that licenses reusing a historical distribution to simulate a new study---plausible when the same class of drugs has been studied in a related setting, as in the motivating scenario, but a genuine assumption that may not hold if the new arm's mechanism reshapes the curve in a way the targets cannot express. Second, we evaluate the method on a single trial rather than through a controlled simulation study, which does not characterize the method's performance across a range of data-generating mechanisms, censoring patterns, and sample sizes, and external replication on additional trials would strengthen the evidence. Third, the upper tail of every simulated curve is governed by a single selected working parametric family, so the method is non-parametric in the body but parametric in the tail, and the tail-weighted information criterion used for family selection is a heuristic. The practical impact of this is limited when the simulated study is subject to administrative censoring, since the same censoring is applied to the synthetic patients and most imputed times in the working parametric tail fall beyond the follow-up horizon, where they are re-censored and do not enter the analysis. Fourth, the quality of a simulated arm depends on the targets the user supplies. Because the method faithfully calibrates the simulated distribution to whatever percentile--time targets it is given, choosing targets that reflect a clinically plausible treatment effect is a matter of domain knowledge; the method itself offers no internal check that the targets are well chosen. Finally, although we have developed the percentile-matching approach for survival endpoints, it is not specific to the survival
setting. The same construction can also be applied to continuous endpoints. In a weight-loss study, for example, the proportion of subjects losing at least 5\%, 10\%, 15\%, 20\%, or 25\% of their body weight is often reported; these thresholds are percentiles of the weight-change distribution and can serve directly as targets, so that a weight-loss arm can be simulated from historical data and this set of response proportions.

In summary, despite the proposed empirical simulation method used some assumptions and approximations based on existing individual patient--level data, it can simulate multivariate data with different types of variables approximately match the real data and requires less assumptions, compared the traditional simulation methods based parametric models. We expect a broad application of this method in future clinical trial design.

\section{Disclosure statement}\label{disclosure-statement}
Authors JS, YD, YL, and YQ are employees and minor shareholders of Eli Lilly and Company. Author YC declares no competing interests.

\section{Data Availability Statement}\label{data-availability-statement}
The data that support the findings of this study are from the OAK trial (ClinicalTrials.gov identifier NCT02008227) that had a tumor mutational burden score defined by \cite{gandara2018blood}. The authors did not generate these data, and restrictions apply to their redistribution. Researchers may request access to individual patient-level data through the clinical study data request platform Vivli (\url{https://vivli.org}). Further details on Roche's criteria for eligible studies are available at \url{https://vivli.org/ourmember/roche/}. The R package \texttt{EmpiricalSim} implementing the proposed methods is provided as Supplementary Material.

\phantomsection\label{supplementary-material}
\bigskip
 
\begin{center}
 
{\large\bf SUPPLEMENTARY MATERIAL}
 
\end{center}
 
\begin{description}
\item[R-package \texttt{EmpiricalSim}:]
The \texttt{EmpiricalSim} R package on https://github.com/ychen-98/EmpiricalSim contains code to perform the simulation methods described in the article.
\end{description}
 
% \section{BibTeX}\label{bibtex}
   
  \bibliography{bibliography.bib}

\end{document}